\documentclass[%
 reprint,
superscriptaddress,
 amsmath,amssymb,
 aps,
]{revtex4-2}

\usepackage{graphicx}
\usepackage{dcolumn}
\usepackage{bm}
\usepackage{xcolor} 
\usepackage{hyperref}
\usepackage{upgreek}

\usepackage[english]{babel}

\begin{document}

\preprint{APS/123-QED}

\title{Robust spin relaxometry with fast adaptive Bayesian estimation}

\author{Michael Caouette-Mansour}
\affiliation{%
Department of Physics, McGill University, 3600 Rue University, Montreal QC, H3A 2T8, Canada
}%
\author{Adrian Solyom }
\affiliation{%
Department of Physics, McGill University, 3600 Rue University, Montreal QC, H3A 2T8, Canada
}%
\author{Brandon Ruffolo }
\affiliation{%
Department of Physics, McGill University, 3600 Rue University, Montreal QC, H3A 2T8, Canada
}%
\author{Robert D. McMichael}
\affiliation{%
National Institute of Standards and Technology, Gaithersburg, MD 20899, USA
}%
\author{Jack Sankey }
\affiliation{%
Department of Physics, McGill University, 3600 Rue University, Montreal QC, H3A 2T8, Canada
}%
\author{Lilian Childress}%
 \email{lilian.childress@mcgill.ca}
\affiliation{%
Department of Physics, McGill University, 3600 Rue University, Montreal QC, H3A 2T8, Canada
}%


\date{\today}

\begin{abstract}

Spin relaxometry with nitrogen-vacancy (NV) centers in diamond offers a spectrally selective, atomically localized, and calibrated measurement of microwave-frequency magnetic noise, presenting a versatile probe for condensed matter and biological systems. Typically, relaxation rates are estimated with curve-fitting techniques that do not provide optimal sensitivity, often leading to long acquisition times that are particularly detrimental in systems prone to drift or other dynamics of interest. Here we show that adaptive Bayesian estimation is well suited to this problem, producing dynamic relaxometry pulse sequences that rapidly find an optimal operating regime. In many situations (including the system we employ), this approach can speed the acquisition by an order of magnitude. We also present a four-signal measurement protocol that is robust to drifts in spin readout contrast, polarization, and microwave pulse fidelity while still achieving near-optimal sensitivity. The combined technique offers a practical, hardware-agnostic approach for a wide range of NV relaxometry applications.

\end{abstract}

\maketitle

\section{Introduction}
Over the last decade, spin sensors based on the nitrogen-vacancy (NV) center in diamond have become a widespread tool for precise, spatially-resolved measurements of local magnetic fields and magnetic noise,
employing a variety of protocols to detect signals from DC to GHz regimes ~\cite{taylor_high-sensitivity_2008, degen_scanning_2008, rondin_magnetometry_2014}. 
One robust and powerful approach to sensing high-frequency magnetic noise is relaxometry, which measures the spin lifetime to detect the magnetic noise spectral density at the spin transition frequency ~\cite{tetienne_spin_2013, steinert_magnetic_2013}. Such measurements are particularly promising probes of biological~\cite{zhang_toward_2021} and condensed-matter systems~\cite{casola_probing_2018}, enabling observations of free radicals in individual mitochondria~\cite{nie_quantum_nodate}, antiferromagnetic textures~\cite{finco_imaging_2021}, electron-phonon interactions in graphene~\cite{andersen_electron-phonon_2019}, spin transfer effects in magnetic insulators \cite{du_control_2017} and / or metals \cite{solyom_probing_2018}, magnon dynamics~\cite{mccullian_broadband_2020} and more.  


A primary challenge in NV relaxometry is the long acquisition time required to estimate the spin decay rate, which limits the space of possible measurements, especially for drift-prone setups or short-lived samples. Slow acquisition is partly due to the low signal-to-noise fluorescence spin readout commonly employed in room-temperature experiments, though advances in high-collection-efficiency nanostructures~\cite{hadden_strongly_2010}, spin-to-charge conversion~\cite{shields_efficient_2015, hopper_near-infrared-assisted_2016, jaskula_improved_2019} and low-temperature detection techniques~\cite{robledo_high-fidelity_2011, zhang_high-fidelity_2021}, offer promising avenues for improvement. In addition to improvements in readout physics, the choice of settings can have a profound effect on measurement efficiency.  Without some prior knowledge of the results, a static list of measurement settings is very likely to spend time on measurements that will have little impact on the results. Especially in high-dynamic-range relaxometry measurements, (i.e. where the initial range of possible decay times is very large compared to the desired precision of the measurement) the need to allow for a wide range of possible results will lead to inefficient non-adaptive protocols.

Bayesian adaptive approaches are well-suited to situations with a high dynamic range. By employing previous outcomes to improve subsequent measurements, adaptive techniques iteratively seek the most sensitive measurement protocol. In the context of NV sensing, adaptive methods have yielded significant speedups in DC magnetometry~\cite{bonato_optimized_2016, dinani_bayesian_2019, dushenko_sequential_2020, mcmichael_sequential_2021}, characterization of nuclear spins~\cite{joas_online_2021}, and charge state detection~\cite{danjou_maximal_2016}. 

In this article, we present a two-pronged approach to overcoming the intertwined problems of long acquisition time and apparatus drift. First, we present a measurement protocol that is insensitive to commonly drifting experimental parameters including fluorescence, contrast, and spin polarization, thereby facilitating reliable data acquisition over longer time scales. Second, we present a Bayesian adaptive measurement technique suitable for NV spin relaxometry, specifically discussing (i) an implementation using a recently developed general framework \cite{mcmichael_optbayesexpt_2021, dushenko_sequential_2020, mcmichael_sequential_2021} and (ii) an approximate implementation requiring fewer computational resources. Importantly, when comparing the adaptive protocols to a ``reasonable'' non-adaptive fitting method, we often find more than an order of magnitude reduction in acquisition time. The adaptive approach is especially well suited to systems with widely varying relaxation rates, and, combined with the drift-insensitive protocol, should enable previously inaccessible avenues of research. Note the approach discussed below should be applicable across many types of spin sensors, but we develop it here using the NV center as a well-developed test-bed system.

\section{Drift-insensitive measurement\label{sec:Drift-indepedant-measurement}}

\begin{figure*}[htb]
\includegraphics[width=15cm]{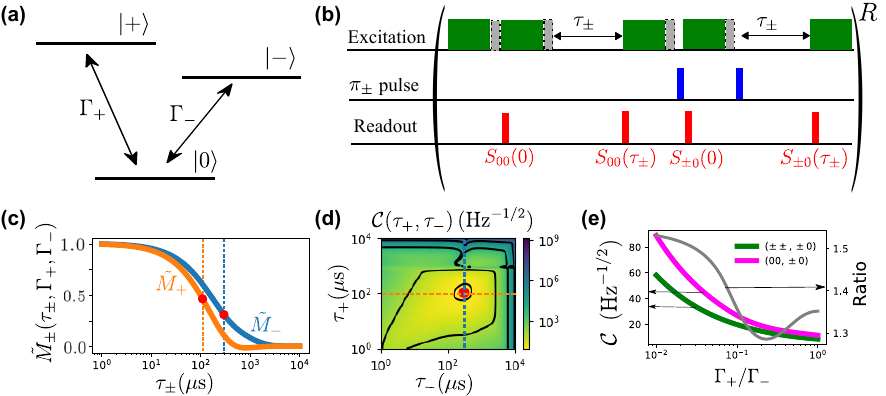}
\caption{
\label{fig:Suggested-measurement-of}
\textbf{Drift-insensitive relaxometry measurements with an NV center.}
(a) Model for NV ground state spin relaxation with decay rates $\Gamma_\pm$. 
(b) Example of NV pulse sequence used to obtain a measurement $M_\pm(\tau_\pm)$ (Eq.~\ref{eq:M}). Optical excitation (green) initializes the spin into $|0\rangle$, and fluorescence photons are counted during readout intervals (red). Microwave $\pi_\pm$ pulses (blue) drive one of the $|0\rangle \leftrightarrow |\pm\rangle$ transitions, and a small additional delay $\approx 1~\upmu$s (grey and dash contour) is inserted after optical pumping to allow singlet deshelving. This sequence is repeated $R$ times, and the protocol is executed twice, exchanging $+ \leftrightarrow -$, to obtain a measurement pair $\{M_+(\tau_+), M_-(\tau_-)\}$.  
(c)  Time dependence of the expected drift-insensitive measurement pair $\tilde{M}_{\pm}(\tau_{\pm},\Gamma_{+},\Gamma_{-})$  for $\Gamma_{+}=3$~ms$^{-1}$ and $\Gamma_{-}=1$~ms$^{-1}$. The dashed lines indicate the optimal delays $\tau_\pm$ (as in (d)).
(d) 
Cost function $\mathcal{C}$ (Eq.~\ref{eq:cost}, with $T_0=0$) evaluated for $\Gamma_{+}=$ 1~ms$^{-1}$ and $\Gamma_{-}=$ 3~ms$^{-1}$, minimized to obtain the optimal $\tau_{+}$ and $\tau_{-}$ (dashed lines).
(e) 
(Left axis) Comparison of cost functions for the optimal measurement pair $(\pm\pm,\pm 0)$ and the drift-insensitive measurement pair $(00,\pm 0)$, for various ratios of the two rates. For each rate, the cost is evaluated at the optimal $\tau_\pm$. (Right axis) Ratio of the optimal measurement sensitivity to the drift-insensitive measurement sensitivity (pink divided by green). }
\end{figure*}

In this section, we outline a simple measurement protocol that is nominally insensitive to commonly drifting experimental parameters such as spin readout contrast, polarization, and microwave pulse fidelity.

As shown in Fig.~\ref{fig:Suggested-measurement-of}(a), the NV electronic ground state is a spin 1 system comprising three Zeeman sublevels denoted $|-\rangle,|0\rangle,|+\rangle$, with magnetic numbers $m_s = -1, 0, 1$, respectively. We assume the $|\pm\rangle$ levels are split by an applied static magnetic field. Optical excitation with green light preferentially populates $|0\rangle$, and $|0\rangle$ produces a larger fluorescence signal than $|\pm \rangle$, providing basic optical preparation and readout of a single spin \cite{childress_atom-like_2014}. In principle, relaxation can occur between any pair of Zeeman sublevels, though classical magnetic noise primarily drives single-quantum transitions, with equal rates of emission and absorption. For NV spins reasonably far from surfaces \cite{myers_double-quantum_2017} and bathed in classical magnetic noise, we model the spin relaxation using two rates $\Gamma_\pm$ that capture bidirectional thermalization between $|0\rangle$ and $|\pm 1\rangle$. Since each rate is proportional to the magnetic noise spectral density at the relevant spin transition frequency \cite{slichter_principles_1990}, measurement of both rates provides information on two spectral components of the environmental magnetic noise. 

Many approaches exist for studying NV decay rates. For example, the decay can be probed all-optically by pumping the spin into $|0\rangle$, waiting for a time $\tau$, and then detecting the population remaining in $|0\rangle$ via fluorescence~\cite{tetienne_spin_2013}; for fast decays, the direct fluorescence under continuous-wave optical excitation can also be used~\cite{rollo_quantitative_2021}. Despite their appealing simplicity, all-optical approaches suffer several drawbacks: most notably, they cannot distinguish $\Gamma_+$ and $\Gamma_-$, and they are sensitive to any time-varying background signals (as well as other drifting quantities such as laser power). Alternately, microwave pulses can be used to prepare and detect populations in different spin sub-levels, often using two measurements to determine both decay rates (e.g., Ref.~\cite{sar_nanometre-scale_2015}). Here we employ such a two-measurement approach, with each measurement comprising four separate ``signals,'' each from a fluorescence detection of differently prepared spins, with spin preparation and detection chosen to eliminate many commonplace experimental challenges. 

One common difficulty is a time-varying background fluorescence, particularly in samples with high enough NV density that it is essentially impossible to avoid weakly exciting neighboring NVs. We can eliminate this background by taking the difference $S_1 - S_2$ between two fluorescence signals $S_j$ that differ only in how the spin is prepared and/or detected. For example, $S_2(\tau)$ might involve preparation into $|0\rangle$, a wait time $\tau$, and a (microwave) $\pi$ pulse driving $|0\rangle \leftrightarrow |+\rangle$ before fluorescence detection (i.e., preparing $|0\rangle$ and detecting population evolution in $|+\rangle$), while $S_1(\tau)$ might lack the $\pi$ pulse (i.e., preparing and detecting in $|0\rangle$). In experiments with large magnetic field gradients, such as those probing ferromagnets, subtracting signals that differ only by the presence of $\pi$-pulses also efficiently isolates the signal from a single NV, since nearby NVs will have sufficiently different transition frequencies that they do not respond to the $\pi$-pulses.

Drifts in optical alignment, optical and microwave excitation power, and DC magnetic fields also affect the measured signals $S_j(\tau)$; such drifts produce shifts in the expected collected fluorescence counts $f_0$ from state $|0\rangle$, the ``contrast'' $C$ defining expected counts $f_\pm \equiv f_0 (1-C)$ from states $|\pm\rangle$, the probability $\alpha$ that the spin is in $|0\rangle$ after optical pumping (i.e., the ``pump fidelity''), and the probability $\eta$ that the $\pi$-pulse does not flip the spin (i.e., the ``$\pi$-pulse error''). These parameters must all either remain constant through an entire (sometimes days-long) experiment or undergo regular re-calibration. A commonly-employed trick (e.g., Ref.~\cite{childress_coherent_2006})  is to normalize the signal by the fluorescence measured after optical pumping, which has an expected value $\alpha f_0 + (1-\alpha)(1-C)f_0$. This eliminates the common prefactor $f_0$, thereby greatly reducing the impact of laser power fluctuations and focus drift; however the result still depends on $C$ and $\alpha$, which themselves depend on laser power. 

Instead, we normalize by the outcome at $\tau=0$, such that our measured quantity is 
\begin{align}\label{eq:M-insensitive}
M(\tau) = \frac{S_1(\tau)-S_2(\tau)}{S_1(0) - S_2(0)}.
\end{align}
A simple rate equation model (Appendices~\ref{app:Dynamical-model-for}-\ref{app:Photoluminescence-to-Signal}) predicts that this ratio is nominally independent of $f_0$, $\alpha$, and $C$, and restricting ourselves to matched $\tau$ also suppresses the impact of time-dependent background fluorescence. Furthermore, since $\tau=0$ measurements are comparatively fast, this normalization adds little overhead, especially in the most sensitive (i.e., long-lifetime) applications. Note that statistical fluctuations can cause the denominator to encompass zero when the signal is too low; in our measurements, we collect sufficient counts to avoid this complicated regime, but remain mindful of the inherent measurement bias associated with ratio estimation (see Appendix~\ref{app:nonlinear-error}). 

We now face a decision of which signals $S_1$ and $S_2$ to employ: by appropriately applying $\pi$-pulses, we can initialize and detect the population in any of the three spin states. 

One approach is to minimize sensitivity to drifts in either microwave power or static magnetic field, which cause drifts in the $\pi$-pulse error $\eta$, making it desirable to pick an $\eta$-independent measurement. Adopting the convention that $S_{ab}(\tau)$ represents preparation in state $|a\rangle$ and readout in state $|b\rangle$ after time $\tau$, only two signal pairs yield an $\eta$-insensitive, robust measurement (see Appendix~\ref{app:Photoluminescence-to-Signal}): $S_1(\tau) \rightarrow S_{00}(\tau)$, $S_2(\tau) \rightarrow S_{+0}(\tau)$ and $S_1(\tau) \rightarrow S_{00}(\tau)$, $S_2(\tau) \rightarrow S_{-0}(\tau)$. 
Combined with Eq.~\ref{eq:M-insensitive}, these yield two drift-insensitive measurements 
\begin{equation} \label{eq:M}
M_{\pm}(\tau_\pm) = \frac{ S_{00}(\tau_\pm) -  S_{\pm 0}(\tau_\pm)}{S_{00}(0) -  S_{\pm 0}(0)},
\end{equation}
where we explicitly allow $\tau_+\ne\tau_-$ to maximize the joint sensitivity to $\Gamma_\pm$ during the adaptive protocols discussed below. Figure~\ref{fig:Suggested-measurement-of}(b) illustrates a pulse sequence (repeated $R$ times to collect sufficient fluorescence) for one of these measurements.

In the limit where the NV photophysics and $\pi$-pulses are negligibly fast compared to the delays $\tau_\pm$, $M_\pm(\tau_\pm)$ directly probes the population dynamics. The rate equation model (Appendix~\ref{app:Photoluminescence-to-Signal}) predicts that, when $M_{\pm}(\tau_\pm)$ is evaluated with the \emph{expected} values $\langle S_{ab}(\tau)\rangle$, we find a ``model function''
\begin{equation}
\tilde{M}_{\pm}(\tau_\pm,\Gamma_+,\Gamma_-)=\frac{(\mathcal{G}+\Gamma_{\pm})e^{-\beta_{+}\tau_{\pm}}+(\mathcal{G}-\Gamma_{\pm})e^{-\beta_{-}\tau_{\pm}}}{2\mathcal{G}},\label{eq:tilde-M-main}
\end{equation}
where $\beta_{\pm}\equiv\Gamma_{+}+\Gamma_{-}\pm\mathcal{G}$ and $\mathcal{G}\equiv\sqrt{\Gamma_{+}^{2}+\Gamma_{-}^{2}-\Gamma_{+}\Gamma_{-}}$, which depends {\it only} on the rates $\Gamma_\pm$ and delays $\tau_\pm$.  Figure \ref{fig:Suggested-measurement-of}(c) shows an example of $\tilde{M}_{\pm}$ for $\Gamma_+ = 1$~ms$^{-1}$ and $\Gamma_- = 3$~ms$^{-1}$; the two measurements evolve with different time scales because $M_+$ ($M_-$) is more sensitive to $\Gamma_+$ $(\Gamma_-)$. As one might expect, the greatest information is gained when probing somewhere in the transition region of these curves. Indeed, for known rates $\Gamma_\pm$, we can identify the optimal delays $\tau_\pm$ (dashed lines) by minimizing the measurement cost (see section~\ref{NOB}) as shown in  Fig.~\ref{fig:Suggested-measurement-of}(d). 
Note that the contours of the cost function in  Fig.~\ref{fig:Suggested-measurement-of}(d) are angled, indicating that $\tau_+$ and $\tau_-$ cannot be selected independently, but must be determined by a 2D optimization of the cost. 

While our proposed ``robust" measurement pair (Eq.~\ref{eq:M}) is minimally affected by drifts in experimental parameters, other choices for $S_1$ and $S_2$ offer improved sensitivity to $\Gamma_\pm$ in experiments where $\eta$ is sufficiently stable. Notably, the ``optimal" measurement pair formed by $S_1 = S_{\pm 0}, S_2 = S_{\pm \pm}$ offers the best rate sensitivity over a wide range of underlying $\Gamma_\pm$ (see Appendix~\ref{app:comparing-protocols} for details). Nevertheless, Figure~\ref{fig:Suggested-measurement-of}(e) shows the optimal ($\eta$-dependent) measurement pair is only 30~\% to 50~\% more sensitive than the robust measurement pair, and we therefore employ the latter in our simulations and experiments. This sensitivity comparison is made assuming that all measurements are made at the optimal $\tau_\pm$ for each protocol, which is feasible for theoretical calculations where $\Gamma_\pm$ are assumed known. Without knowing $\Gamma_\pm$, however, we cannot know the optimal $\tau_\pm$, motivating the adaptive protocols discussed below.

\begin{figure*}
\includegraphics[width=13cm]{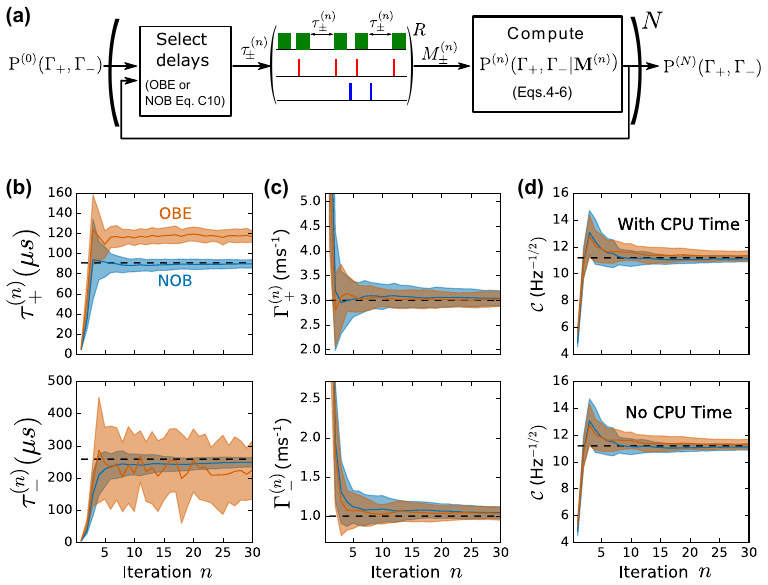}
\caption{
\label{fig2:adaptive-approach} 
\textbf{Monte Carlo simulation of the algorithms} 
(a) Adaptive protocol for measuring $\Gamma_{\pm}$. The prior $\mbox{P}^{(0)}(\Gamma_+, \Gamma_-)$ or most recent distribution $\mbox{P}^{(n-1)}(\Gamma_+, \Gamma_-| \mbox{\textbf{M}}^{(n-1)})$ is used to calculate optimal delays $\tau_\pm^{(n)}$ (using OBE or NOB), and the posterior distribution $\text{P}^{(n)}(\Gamma_+,\Gamma_-,\mathbf{M}^{(n)})$ is inferred after each acquisition. This loop is repeated $N$ times. 
(b) Optimal delays $\tau_{\pm}$ of each iteration determined by OBE (orange lines) and NOB (blue lines). Dashed lines are value minimizing cost $\mathcal{C}$ (Eq.~\ref{eq:cost}). 
(c) Estimated rates after each iteration. The dashed lines indicate the true values of the simulation. 
(d) Evolution of $\mathcal{C}$ with (top) and without (bottom) CPU time taken to find the optimal $\tau_\pm$. 
In this simulation, the CPU overhead per iteration was 2 s for OBE and 0.3 s for NOB. The simulation parameters are: $\Gamma_+ = 3$~ms$^{-1}$, $\Gamma_- = 1$~ms$^{-1}$, average photon counts per detection $f_0 = 0.02$, fluorescence contrast $C=0.24$, background fluorescence $b(\tau)=0$, polarization $\alpha = 0.8$, $\pi$ pulse errors $\eta_\pm = 0.05$, number of readouts $R=10^6$. The filled band indicates one standard deviation and the solid line indicates the mean over 30 simulations. }
\end{figure*}

\section{Bayesian Adaptive protocol}

In this section we introduce Bayesian inference within the context of spin-1 relaxometry, focusing on how to select optimal delays $\tau_\pm$ using two numerical methods. We note that both of these methods could be adapted to any such two-measurement approach, and simpler versions could be applied to all-optical methods.

\subsection{Bayesian formalism for spin-1 relaxometry}

In order to maximize the efficacy of each measurement, we employ the Bayesian adaptive approach illustrated schematically in Fig.~\ref{fig2:adaptive-approach}(a). This is used to iteratively improve upon a 2D probability density function (PDF) $\text{P}(\Gamma_+, \Gamma_-)$ that describes our knowledge of $\Gamma_\pm$. In each ($n^\text{th}$) iteration of the loop, we use the previous PDF  $\text{P}^{(n-1)}(\Gamma_+, \Gamma_- | \mathbf{M}^{(n-1)})$, inferred from all $n-1$ previous measurements $\mathbf{M}^{(n-1)} = \{M^{(n-1)}_+, M^{(n-1)}_-, ... , M^{(1)}_+, M^{(1)}_-\}$, to select optimal delays $\tau^{(n)}_\pm$ for the next measurement. We then perform the measurement and use Bayes theorem to find the $n^\text{th}$ PDF
\begin{equation}\label{eq:bayesian-update}
\begin{split}
\text{P}^{(n)}&(\Gamma_+, \Gamma_-|\mathbf{M}^{(n)}) \propto \\
&\tilde{\text{P}}^{(n)}(M^{(n)}_+,M^{(n)}_-|\Gamma_+,\Gamma_-)\text{P}^{(n-1)}(\Gamma_+,\Gamma_-|\mathbf{M}^{(n-1)}),
\end{split}
\end{equation}
where
\begin{align}
    \label{eq:Likelihood_signals}\tilde{\text{P}}^{(n)}(M^{(n)}_+,M^{(n)}_-|\Gamma_+,\Gamma_-) \propto e^{-\left(\chi^{(n)}_{+}\right)^{2}-\left(\chi^{(n)}_{-}\right)^{2}} \\
\chi^{(n)}_{\pm} \equiv \frac{M^{(n)}_\pm-\tilde{M}_{\pm}(\tau_{\pm}^{(n)},\Gamma_{+},\Gamma_{-})}{\sqrt{2}\sigma_{M_{\pm}}^{(n)}}\label{eq:chi_signals_app}
\end{align}
is the likelihood, i.e., the modeled probability of obtaining the measurement data $M_\pm^{(n)}$ at delay $\tau_\pm$ given underlying rates $\Gamma_\pm$. The likelihood given in Eq.~\ref{eq:Likelihood_signals}-\ref{eq:chi_signals_app} is derived assuming sufficiently high signal to noise that the measurement outcomes are approximately Gaussian distributed. The measurement outcomes $M_\pm^{(n)}$ and uncertainties $\sigma_{M_\pm}^{(n)}$ should be calculated with care, 
since the difference in the denominator of Eq.~\ref{eq:M} often leads to nonlinearity that precludes standard error propagation, even when the signals $S_{ab}$ are large enough to assume they are drawn from Gaussian distributions; Appendix \ref{app:nonlinear-error} provides further discussion and a somewhat improved approximation. 

To initialize the procedure, we assume a uniform distribution $\text{P}^{(0)}$ spanning all ``reasonable'' values of $\Gamma_+$ and $\Gamma_-$. The measurement and update process is repeated $N$ times, and we numerically normalize $\mathrm{P}^{(n)}$ at each step, allowing us to inspect mean values $\bar{\Gamma}_\pm$ and uncertainties $\sigma_{\Gamma_\pm}$ at each step. 

In the rest of this section, we introduce two methods for selecting delays $\tau_\pm$: a slightly modified version of the general-purpose library recently developed by NIST \cite{mcmichael_optbayesexpt_2021}, and a near-optimal approximation with significantly reduced computational overhead. Both of these approaches rely on the parametric model (Eq.~\ref{eq:tilde-M-main}), and although they settle on somewhat different ``optimal'' delays, they converge with similar sensitivities. Furthermore, they employ the same Bayesian inference, such that the choice of delay does not bias the estimate of $\Gamma_\pm$. 

\subsection{OptBayesExpt (OBE)}

A detailed description of the OBE library can be found in Ref.~\cite{mcmichael_optbayesexpt_2021}, but we provide a functional description here, along with a modification that accounts for variable measurement time during optimization. Nominally, OBE requires only that we supply a model function $\tilde{M}_\pm$ (Eq.~\ref{eq:tilde-M-main}) and a prior distribution $\text{P}^{(0)}(\Gamma_+,\Gamma_-)$, then feed measurement pairs $M^{(n)}_\pm$ at each iteration of the loop in Fig.~\ref{fig2:adaptive-approach}(a), allowing it to estimate the optimal delays and update the distribution. In this algorithm, the distributions $\text{P}^{(n)}$ are represented by particle filters, stochastic clouds of points, each with $(\Gamma_+, \Gamma_-)$ coordinates and a weight value. 
For the prior $\text{P}^{(0)}$, we supply an evenly spaced grid spanning all ``reasonable'' values with each point having the same weight. 

In order to select delays $\tau_\pm^{(n)}$, OBE maximizes a utility function quantifying the expected knowledge gained from the next measurement, using the $P^{(n-1)}(\Gamma_+,\Gamma_-)$ distribution as an input. By default, the utility function does not penalize long acquisition times; since we are interested in optimizing sensitivity defined in the usual way (with units of the measured quantity per $\sqrt{\text{Hz}}$; see, e.g., Ref.~\cite{degen_quantum_2017}), we scale the utility function by $1/\sqrt{T}$, where $T \approx 2R\left(\tau_++\tau_-\right)+T_0$ is the total acquisition time, and $T_0$ includes any relevant overhead time (and may depend on $R$). 

The orange curves in Fig.~\ref{fig2:adaptive-approach}(b)-(c) show how the selected $\tau_\pm^{(n)}$ and estimated $\Gamma_\pm^{(n)}$ converge with each iteration applied to typical simulated data. Here we assume ``true'' rates $\Gamma_+^\text{true}=3$~ms$^{-1}$ and $\Gamma_-^\text{true}=1$~ms$^{-1}$ (dashed lines in (c)), use a number $\mathcal{N}=10^5$ points to define the cloud at each step, and let $\text{P}^{(0)}$ span 0.055~ms$^{-1}\le \Gamma_\pm \le$ 100~ms$^{-1}$ (these bounds are motivated in the experimental section below) while allowed delay times lie on a 1000 $\times$ 1000 point, logarithmically spaced grid spanning $3~\upmu\text{s} < \tau_\pm^{(n)} < 5.5~\text{ms}$ (i.e., the optimal times associated with the bounds on $\Gamma_\pm$). We simulate measurement outcomes using signals with a Poissonian distribution around the expected fluorescence counts given by our rate equation model (Appendix~\ref{app:Photoluminescence-to-Signal}), with parameters as given in the Fig.~\ref{fig2:adaptive-approach} caption.
This example shows how the stochastic nature of this algorithm can cause large fluctuations in probe time; in this system, we find the algorithm has difficulty converging for $\mathcal{N}\lesssim 10^4$, while larger values of $\mathcal{N}$ introduce significant computational overhead. Importantly, despite the large fluctuations in $\tau_-^{(n)}$ shown in this example, the estimated values of $\Gamma_\pm$ still rapidly converge without bias. 

\subsection{Near-optimal Bayes (NOB)}
\label{NOB}

Selecting the optimal $\tau^{(n)}_\pm$ represents a dominant computational step in the aforementioned adaptive loop. In this section, we present an alternative ``near-optimal Bayes'' (NOB) approach that bypasses most of this overhead. Ultimately, we recommend trying OBE until a computational bottleneck arises, then falling back on NOB. As discussed in Sec.~\ref{sec:comparison}, despite discarding information, this approach converges upon the correct values of $\Gamma_\pm$, with a sensitivity comparable to that of OBE.

To select optimal delays $\tau_\pm^{(n)}$, we seek a ``cost'' or ``inverse utility'' function $\mathcal{C}(\tau_+,\tau_-)$ that serves as a proxy for the expected sensitivity given a prior distribution and choice of $\tau_\pm^{(n)}$. However, since we are estimating \textit{two} quantities, each having its own sensitivity, there is some freedom in the definition of $\mathcal{C}$ that, ultimately, depends on the goal of the experiment. In our case, a reasonable choice is the combined fractional sensitivities
\begin{equation}
\mathcal{C}(\tau_+,\tau_-)=\sqrt{\left(\frac{\sigma_{\Gamma_{+}}}{\Gamma_{+}}\right)^2 +\left(\frac{\sigma_{\Gamma_{-}}}{\Gamma_{-}}\right)^2}\sqrt{T}.\label{eq:cost}
\end{equation}
Nominally, $\Gamma_\pm$ and $\sigma_{\Gamma_\pm}$ should be calculated from (i) our most up-to-date distribution $\textrm{P}^{(n-1)}(\Gamma_+,\Gamma_-|\mathbf{M}^{(n-1)})$ and (ii) the distribution $\tilde{\text{P}}^{(n)}(M^{(n)}_+,M^{(n)}_-|\Gamma_+,\Gamma_-)$ of expected measurements for each possible $\Gamma_\pm$ (Eq.~\ref{eq:Likelihood_signals}), which yield 
the resulting Bayes-updated distribution $\textrm{P}^{(n)}(\Gamma_+,\Gamma_-|\mathbf{M}^{(n)})$ for each of these possible measurement outcomes (where the $n^\text{th}$ measurement is hypothetical). The difficulty of manually checking every possibility afforded by these 
distributions motivates the Monte Carlo approach of OBE.

In the NOB approach, we circumvent the issue by constructing an estimate of the cost function that can be minimized with respect to $\tau_\pm$ without having to calculate $\textrm{P}^{(n)}(\Gamma_+,\Gamma_-|\mathbf{M}^{(n)})$ for every possible measurement delay. Essentially, we suppose that the true underlying rates are given by the mean values $\bar{\Gamma}^{(n-1)}_\pm$ of the prior $\textrm{P}^{(n-1)}(\Gamma_+, \Gamma_-)$. We then evaluate $\mathcal{C}$ for a single measurement (i.e. with a uniform prior) on a system with those underlying rates. Note that this approximation throws out the details of previous measurements and potential interplay between previous and future measurements, reducing our current knowledge of the system to the mean values of the rate distribution. 

Mathematically, the NOB approach estimates $\mathcal{C}$ by finding the rate uncertainties $\sigma_{\Gamma_\pm}$ obtained from a single measurement pair performed at delays $\tau_\pm$ on a system with known underlying rates, and thus known $\tilde{M}_\pm(\tau_\pm^{(n)})$ (Eq.~\ref{eq:tilde-M-main}). We assume that the measurement outcomes $M_\pm$ are Gaussian distributed with means $\tilde{M}_\pm(\tau_\pm^{(n)})$ and variances $\sigma_{M_\pm}^2$, as appropriate to high SNR measurements; we then approximate the resulting posterior distribution for $\Gamma_\pm$ as Gaussian, which allows us to relate the rate uncertainties $\sigma_{\Gamma_\pm}(\tau_\pm)$ to the measurement uncertainties $\sigma_{M_\pm}$ via derivatives of the form $\partial\tilde{M}_\pm(\tau_\pm^{(n)})/\partial \Gamma_\pm$ (see details in Appendix \ref{app:calculating-cost}). Ultimately, we obtain an analytic (if involved) estimate for $\mathcal{C}$ as a function of $T$, $\tau_\pm$, $\sigma_{M_\pm}$, and assumed ``true" rates $\Gamma_\pm$. 

At this point, our estimate for $\mathcal{C}$ involves measurement uncertainties  $\sigma_{M_\pm}$ that can be evaluated by modeling our signals (as in Appendix~\ref{app:Photoluminescence-to-Signal}) and propagating their shot noise through to the measurement (see Appendix~\ref{app:nonlinear-error}). In general the resulting expressions for  $\sigma_{M_-}$ and $\sigma_{M_+}$ are different and delay-dependent, but examination of Eq.~\ref{eq:M} suggests that for low-contrast fluorescence signals, the variation is likely to be small. Thus, to yet further speed up our minimization of $\mathcal{C}$ with respect to $\tau_\pm$, we approximate both $\sigma_{M_+}$ and $\sigma_{M_-}$ as the same constant $\sigma_M$, which then factors out of our expression for $C$. The end result is an analytic function of $T, \tau_\pm$, and $\Gamma_\pm$ (see Eq.~\ref{eq:approxC}) that can be minimized to find near-optimal delays.   

 In practice, we calculate $\mathcal{C}$ on a 1000 $\times$ 1000 point, logarithmically spaced grid spanning $3~\upmu\text{s} < \tau_\pm^{(n)} < 5.5~\text{ms}$, as used for OBE. To further reduce overhead, we represent each distribution $\mathrm{P}^{(n)}(\Gamma_+,\Gamma_-)$ with an evenly spaced 200 $\times$ 200 grid of weights spanning $\pm 10\sigma^{(n-1)}_{\Gamma_\pm}$ around the mean value $\bar{\Gamma}^{(n-1)}_\pm$ (but never exceeding the original bounds), and linearly interpolate the prior as needed.

The blue curves in Fig.~\ref{fig2:adaptive-approach}(b)-(c) show how the NOB-chosen delays $\tau_\pm^{(n)}$ and inferred values $\Gamma_\pm$ evolve with each iteration, using the same system parameters and initial prior as for OBE. This typical example shows how the optimization criteria of OBE and NOB lead to somewhat different values of $\tau_\pm$, but, as shown in Fig.~\ref{fig2:adaptive-approach}d, this does not significantly affect the sensitivity, even if we ignore the added overhead associated with OBE. We also emphasize that both OBE's and NOB's Bayesian inference remains rigorous and does not bias the estimated values of $\Gamma_\pm$.

\section{Comparison to a non-adaptive approach}\label{sec:comparison}

To verify real-world performance of the adaptive protocols, we apply them to relaxometry measurements in the presence of noise from a nearby thin-film magnetic micro-structure similar to that of Ref.~\cite{solyom_probing_2018}. 

\subsection{Device geometry} 
\begin{figure}
\includegraphics[scale=1]{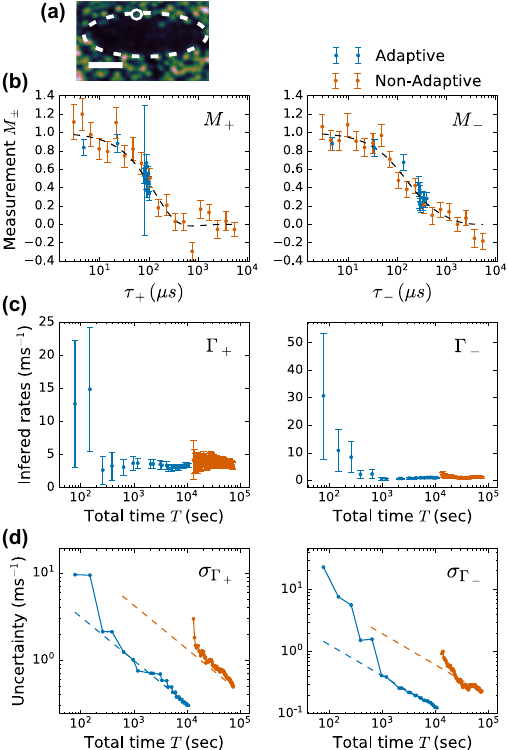}
\caption{
\label{fig3:experimental_comparison}
\textbf{Experimental gain in sensitivity.}
(a) Confocal image of the nanomagnet (dashed contour) and nearby NV fluorescence (light spots). The NV used for this measurement is circled. The white bar corresponds to 2 $\upmu$m. (b) Measurements $M_+(\tau_+)$ (left) and $M_-(\tau_-)$ (right)
for adaptive (NOB) and non-adaptive protocols, with 28~mT applied along the NV axis (open symbols in Fig.~\ref{fig4:field_sweep}). Error bars denote 1 standard deviation propagated from the shot noise of signals.
Bayesian inference yields  
 $\Gamma_{+}$=3.4(3)~ms$^{-1}$, $\Gamma_{-}$ = 1.0(1)~ms$^{-1}$ (adaptive) and 
$\Gamma_{+}$ =3.3(4)~ms$^{-1}$, 
$\Gamma_{-}$=1.2(2)~ms$^{-1}$ (non-adaptive). The dashed lines are the model (Eq.~\ref{eq:tilde-M-main}) using $\Gamma_\pm$ from adaptive data. 
(c) Evolution of $\Gamma_{\pm}$ estimates (error bars show standard deviation) as a function of total elapsed time, including overhead (NOB: $T_0 \approx 85$ s per measurement and $T_0/T \approx 0.20$; NAP: $T_0/T \approx 0.11$). 
Each adaptive point corresponds to a single iteration of the protocol, and each non-adaptive point comes from analysis after each delay is briefly probed.
(d) Evolution of rate uncertainty for the data in (b). 
Dashed lines show $\frac{1}{\sqrt{T}}$ trend from the final data point. }
\end{figure}

Fig.~\ref{fig3:experimental_comparison}(a) shows a fluorescence map near the flat, ellipsoidal, Pt-capped metallic magnet (5~nm~Py / 5~nm~Pt films with lateral dimensions $2~\upmu$m $\times~7~\upmu$m, within the dotted line) and the probed NV (circle), which is implanted approximately 75 nm into the bulk diamond substrate. The stray field gradient from the magnetic layer shifts the NV spin transition frequencies relative to those of the surrounding NVs, allowing us to better isolate a single NV, and the thermal magnetic noise substantially affects the NV spin decay rates. We excite the NV at 532 nm while detecting broadband phonon-sideband fluorescence in a confocal apparatus, and intermittently automatedly re-optimize the focus on the NV; we apply microwave $\pi$ pulses to the NV via a proximal stripline also fabricated on the diamond. A nearby, mm-scale Nd permanent magnet is used to generate static fields from 15~mT to 50 mT along the NV axis.

\subsection{Observed speedup}
\label{subsec:observed-speedup}
We compare the performance of the adaptive protocol to that of a common non-adaptive protocol (NAP), wherein the model function (Eq.~\ref{eq:tilde-M-main}) is fit to measurements from a logarithmically spaced set of delays $\tau_\pm$  (see, e.g., Refs.~\cite{sar_nanometre-scale_2015, du_control_2017}). Here we employ NOB out of convenience; OBE produces similar results.

In the NAP case, the choice of delay times $\tau_{\pm}$ significantly impacts the sensitivity. Motivated by the bounds on T$_1$ times observed in literature \cite{jarmola_temperature-_2012,sar_nanometre-scale_2015}, we choose 20 points logarithmically spaced between 3 $\upmu$s and 5.5~ms; this range is guaranteed to include near-optimal $\tau_\pm$ for room-temperature NVs in both noisy ($\Gamma_\pm\approx$ 100~ms$^{-1}$, $\tau_\pm\approx$ 3 $\upmu$s) and quiet ($\Gamma_\pm\approx$ 0.055~ms$^{-1}$, $\tau_\pm\approx $ 5.5~ms) conditions. 

Figure~\ref{fig3:experimental_comparison} shows a comparison of the NOB and NAP, measured under identical conditions (see caption). Figure~\ref{fig3:experimental_comparison}(b) shows measurements for $\tau_\pm$ chosen by NAP (orange) and NOB (blue), along with the evolution expected from the final estimate of $\Gamma_\pm$ (dashed lines). As is sensible, NOB spends the majority of its measurement time in the transition region, where more information is acquired. The large-error data point likely corresponds to a significant drift in optical alignment, which decreases the SNR of the signals but does not systematically affect the measurement outcome; it has minimal impact on the data analysis and adaptive acquisition. Figure~\ref{fig3:experimental_comparison}(c) shows how the inferred rates $\Gamma_\pm$ evolve as a function of the total experimental time $T$, updated after each NOB iteration or after each set of repetitions of the NAP sequence adds more counts for each $\tau_\pm$. In both approaches, we use the distribution calculated by Bayesian update (Eq.~\ref{eq:bayesian-update}) to estimate mean values and uncertainties. Both converge to the same value within uncertainty, and NOB produces higher precision, as expected. 

To further quantify the speedup, Fig.~\ref{fig3:experimental_comparison}(d) shows the evolution of the uncertainties $\sigma_{\Gamma_\pm}$ (dots), along with the  $1/\sqrt{T}$ trend expected for Gaussian distributions (dashed lines), anchored to the final point, as a guide to the eye. For this example, the NOB uncertainties are reached 18 times faster for $\Gamma_+$ and 23 times faster for $\Gamma_-$. Note the total time $T$ includes everything associated with the measurement: time spent optimizing the laser focus, computation time, and all other dwell times, some of which are in principle avoidable. If we only include the delays $\tau_\pm$, the 81.7~\% (89.5~\%)  duty cycle of NOB (NAP) slightly increases the speedup to 20$\times$ (25$\times$) for $\Gamma_+$ ($\Gamma_-$). 

\begin{figure}[htb]
\includegraphics[width=7cm]{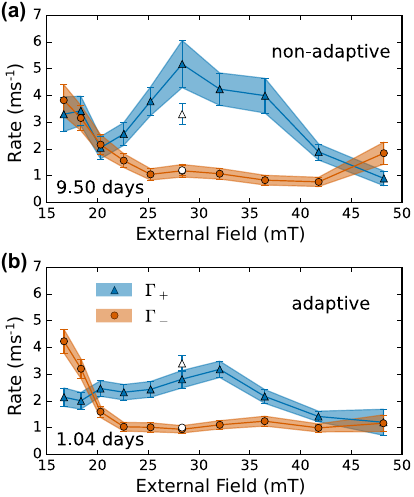}
\caption{\label{fig4:field_sweep}
\textbf{Experimental field dependence of rates.}
Environmental noise is varied by continuously tuning the external DC applied field along the NV axis, with total elapsed time indicated in days. 
(a) Rates obtained using 20-point non-adaptive approach as in Fig.~\ref{fig3:experimental_comparison}.  
(b) Rates obtained with NOB protocol. In both (a) and (b), the open points (also the data explored in Fig.~\ref{fig3:experimental_comparison}) indicate the rates obtained by immediately acquiring with NAP after NOB. }
\end{figure}

An essential motivation for adaptive protocols is their ability to extract rate parameters over a large dynamic range. 
To explore this, we vary the applied field, which shifts the NV transitions and magnetic noise spectrum, thereby changing $\Gamma_\pm$ as shown in Fig.~\ref{fig4:field_sweep} \cite{solyom_probing_2018, du_control_2017}. The advantage provided by the adaptive protocol is striking.  The non-adaptive protocol required 9.5~d while the adaptive protocol required only 1.05~d. 

Comparison of results shown in Fig.~\ref{fig4:field_sweep}(a) and (b) shows  that the estimated values of $\Gamma_\pm$ do not agree well at all fields. 
We suspect this is due to thermally activated reconfiguration of the proximal micro-magnet. At 28 mT, for example, the original two data sets (solid symbols) were taken approximately 10 days apart. Re-taking these points in ``quick'' succession (open symbols and Fig.~\ref{fig3:experimental_comparison}; 3-hour NOB followed by a 21-hour NAP) yields agreement within uncertainty, further motivating the need for the faster acquisition possible with adaptive protocols.

\subsection{Expected speedup for other rates}\label{sec:other_rates}
\begin{figure}
\includegraphics[width=8.3cm]{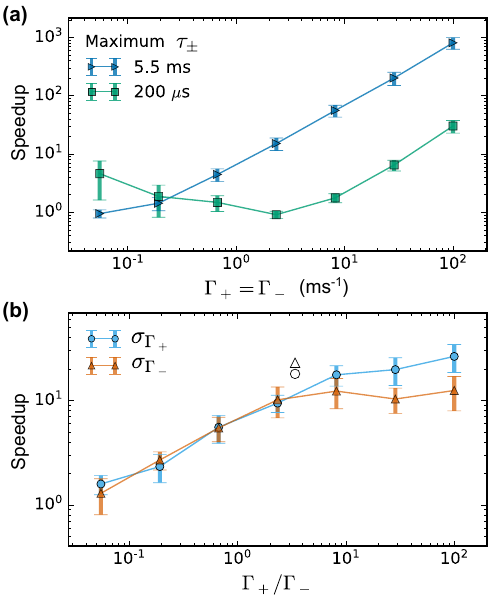}
\caption{\label{fig:montecarlo} 
\textbf{Monte Carlo simulation of speed-up.}
(a) Adaptive speedup relative to non-adaptive, with equal underlying rates and the same overhead as the data in Fig.~\ref{fig3:experimental_comparison} (duty cycle of 80.2~\% for the adaptive and 89.5~\% for the non-adaptive), using 20 logarithmically-spaced NAP delays 3 $\upmu$s $<\tau_\pm < $ 5.5~ms (blue) or 3 $\upmu$s $<\tau_\pm < $ 200 $\upmu$s (green). The results for $\sigma_{\Gamma_\pm}$ were similar (as expected by symmetry), and so only the speedup for $\Gamma_+$ is shown.
(b) Average speedup for $\sigma_{\Gamma_+}$ (blue) and $\sigma_{\Gamma_-}$ (orange) when the underlying rates are different, with 20 logarithmically-spaced non-adaptive delays spanning 3 $\upmu$s $<\tau_\pm <$ 5.5~ms. The open markers are experimental results from Fig.~\ref{fig3:experimental_comparison}.
In both (a) and (b), the solid points indicate the average results over 30 (adaptive) x 30 (non-adaptive) = 900 pairs of simulations, with the error bars representing the standard deviation.
}
\end{figure}
In the measurements leading to Fig.~\ref{fig4:field_sweep}, a trend linking fast adaptive runs with larger relaxation rates was noted. We tentatively explain this trend noting that when rates are high, relaxation completes quickly, meaning that measurements investing long waiting times will yield little information. As shown in Fig.~\ref{fig3:experimental_comparison}(b), the adaptive methods focus measurement times close to the half-life of the decay, so overall, we expect less total waiting time for high decay rates. 

To illustrate this dependence of speedup on underlying relaxation rate and provide a rough idea of what to expect outside of our experimentally accessible rates, we used the same approach and parameters as in Fig.~\ref{fig2:adaptive-approach} to numerically simulate signals over a wide range of relaxation rates. We employed the same protocol as in the experiments, except that we allowed the \emph{adaptive} delays to span a larger range, $1~\upmu\text{s} < \tau_\pm < 1$~s; this has no significant impact on measurements of fast relaxation, but does allow the adaptive protocol to optimally measure an even wider range of $\Gamma_\pm$. 

Figure~\ref{fig:montecarlo} shows the mean speedups obtained from 30 simulations of both NAP and NOB, comparing all $30\times 30 = 900$ possible combinations of outcomes; this was performed for each set of $\Gamma_\pm$, ranging from $0.05$~ms$^{-1}$ (similar to the slowest expected rates for room temperature NVs) to $100$~ms$^{-1}$ (similar to the highest value). Figure~\ref{fig:montecarlo}(a) shows results for equal values of $\Gamma_\pm$, for two different ranges of non-adaptive $\tau_\pm$'s. Intuitively, when the maximum $\tau_\pm$ is 200 $\upmu$s (green) -- a nearly optimal value for $\Gamma_\pm\approx 2$~ms$^{-1}$ -- NOB only has a significant advantage when $\Gamma_\pm$ are far from 2~ms$^{-1}$. Similarly, the advantage vanishes near $\Gamma_\pm$ = 55 s$^{-1}$ when the NAP's maximum $\tau_\pm$ is 5.5~ms. Figure \ref{fig:montecarlo}(b) shows the speedup for different relative rates (with $\Gamma_-$ fixed at 1~ms$^{-1}$). Here we see the advantage is reduced even when only one of the rates becomes small, since at least one of the optimal delays becomes long. For larger $\Gamma_+$, the advantage saturates for the same reason. The open symbols in (b) show the experimentally observed speedup, which is somewhat higher than expected. This result is not at all atypical, and we believe it is due to the aforementioned drift in the underlying rates over the longer NAP times.

\section{Conclusions}

We presented an approach to NV relaxometry incorporating drift-insensitive measurements and adaptive Bayesian estimation. Compared to a ``reasonable'' curve fitting protocol, the adaptive approach essentially always outperforms, often achieving considerable speedup. 

These results come with some notable caveats. First, we emphasize that our choice of non-adaptive approach is based on methods that are neither optimal nor universal. Spacing the delays closer to the optimal values will certainly result in improved sensitivity. However, without prior knowledge of the rates, it would be impractical to optimize the non-adaptive approach in this manner. For this reason, we see no reason not to employ a Bayesian adaptive protocol. Similarly (as noted above), the speedup of a given measurement also depends on the underlying rates. Second, our claim of drift insensitivity is valid only within the presented rate equation approximation. We assume that the microsecond timescale for the photo-physics and NV spin dephasing is negligible compared to the relaxation times we wish to measure; while appropriate to the millisecond rates observed here, this approximation may not be met by NV defects in noisier environments. Additionally, $\pi$-pulse errors are modeled assuming no cross-talk between NV transitions, which can break down especially when working in low magnetic fields.

Nevertheless, our results indicate that adaptive data acquisition techniques offer substantial improvements for relaxometry without any need to change any experimental hardware. In our setup, adaptive protocols were implemented using a standard desktop workstation, without any real-time computation or graphics processor enhancements. Moreover, adaptive acquisition can be readily combined with improvements in spin readout fidelity, e.g., high optical collection efficiency device geometries \cite{hadden_strongly_2010}, spin-to-charge conversion \cite{shields_efficient_2015} or resonant readout \cite{robledo_high-fidelity_2011, zhang_high-fidelity_2021}. The possibility to dramatically reduce acquisition times for relaxometry opens a window on faster magnetic processes in condensed-matter and biological environments, potentially uncovering new phenomena in detection of noise transients. Ultimately, these results illustrate the power of adaptive protocols to achieve high sensitivity while maintaining a large dynamic range.

\section{Acknowledgements}
This work was partially supported by the National Science and Engineering Research Council (NSERC RGPIN 435554-13,  RGPIN-2020-04095), Canada Research Chairs (229003 and 231949), Fonds de Recherche - Nature et Technologies (FRQNT PR-181274), the Canada Foundation for Innovation (Innovation Fund 2015 project 33488 and LOF/CRC 229003), and l'Institut Transdisciplinaire d'Information Quantique (INTRIQ). L. Childress is a CIFAR fellow in the Quantum Information Science program. 
A. Solyom acknowledges support from the NSERC CREATE program QSciTech. 

\section{Author contributions}
M.C-M. developed the protocols, performed the simulations and experiments, analyzed data, and co-wrote the manuscript. A.S. assisted with data analysis and fabrication. B.R. led fabrication of the sample. R.M. assisted with implementation of the OBE protocol. J.S. advised on experiments and simulations and co-wrote the manuscript. L.C. conceived of the approach and the measurement technique, advised on experiments and simulations, and co-wrote the manuscript. 

\bibliography{magnetometry4}

\appendix

\section{Rate equation model\label{app:Dynamical-model-for}}
Here we describe the population dynamics of the three spin sublevels $\{|-\rangle, |0\rangle, |+\rangle\}$ with rate equations
\begin{equation}
\left(\begin{array}{c}
\dot{\rho}_{-}(t)\\
\dot{\rho}_{0}(t)\\
\dot{\rho}_{+}(t)
\end{array}\right)=\left(\begin{array}{ccc}
-\Gamma_{-} & \Gamma_{-} & 0\\
\Gamma_{-} & -(\Gamma_{-}+\Gamma_{+}) & \Gamma_{+}\\
0 & \Gamma_{+} & -\Gamma_{+}
\end{array}\right)\left(\begin{array}{c}
\rho_{-}(t)\\
\rho_{0}(t)\\
\rho_{+}(t)
\end{array}\right)\label{eq:matrix_pop_decay},
\end{equation}
where $\Gamma_{\pm}$ are the decay rates in Fig.~\ref{fig:Suggested-measurement-of}(a), and $\rho_{j}(t)$ is the population in state
$|j\rangle$. These equations can be solved analytically for any initial spin state, yielding solutions we write using the notation that $p_{ij}(\tau)$ is the population in state $|j\rangle$ a time $\tau$ after preparing state $|i\rangle$. If we imagine ``ideal'' signals $S_{ij}(\tau)$ that are proportional to $p_{ij}(\tau)$ with no fluctuations, the measurement scheme of Eq.~\ref{eq:M} yields the ``model function''
\begin{align}
\tilde{M}_\pm(\tau_\pm,\Gamma_+,\Gamma_-) = \frac{p_{00}(\tau_\pm) - p_{\pm 0}(\tau_\pm)}{p_{00}(0) - p_{\pm 0}(0)} ~~~~~\\ ~~~~=\frac{1}{2\mathcal{G}}\left[(\mathcal{G} + \Gamma_{\pm}) e^{-\beta_{+}\tau_\pm}+  (\mathcal{G}-\Gamma_{\pm})e^{-\beta_{-}\tau_\pm}\right],\label{eq:tilde-M-app}
\end{align}
found in Eq.~\ref{eq:tilde-M-main}, where $\beta_{\pm}=\Gamma_{+}+\Gamma_{-}\pm\mathcal{G}$ and $\mathcal{G}=\sqrt{\Gamma_{+}^{2}+\Gamma_{-}^{2}-\Gamma_{+}\Gamma_{-}}$.

\section{Measurement model \label{app:Photoluminescence-to-Signal}}

In reality, the measured signals $S_{ij}$ do not directly probe the expected probabilities $p_{ij}$ due to imperfections in initialization, manipulation, and readout. To include these effects, we calculate the expected average photo-counts
\begin{equation}
\overline{ S_{ij}(\tau)} = R \left[ \mathbf{c}\cdot\mathbf{B}[j]\cdot\mathbf{P}(\tau)\cdot\mathbf{B}[i]\cdot\mathbf{s} + b(\tau) \right] \label{eq:Mmodel}
\end{equation}
after $R$ readouts. In this expression,
\begin{align}
    \mathbf{s}=\left(\begin{array}{c}
(1-\alpha)/2\\
\alpha\\
(1-\alpha)/2
\end{array}\right)
\end{align}
encodes the initial spin state, where $\alpha$ is the initial spin polarization into $|0\rangle$ ($\alpha = 1$ for perfect spin initialization). The $\mathbf{B}[i]$ operator transforms  $|0\rangle \leftrightarrow |i\rangle$.
\begin{align}
\mathbf{B[-]} &= \left(\begin{array}{clcr}
\eta_{-} & 1-\eta_{-} & 0\\
1-\eta_{-} & \eta_- & 0\\
0 & 0 & 1 \\
\end{array}\right) \\
\mathbf{B}[0] &= \left(\begin{array}{clcr}
1 & 0 & 0 \\
0& 1 & 0 \\
0 & 0 & 1 \\
\end{array}\right)&\\
\mathbf{B[+]} &= \left(\begin{array}{clcr}
1 & 0 & 0 \\
0 & \eta_{+} & 1-\eta_{+} \\
0 & 1-\eta_{+} & \eta_+ \\
\end{array}\right),
\end{align}
where $\mathbf{B}[\pm]$ are $\pi$ pulses and $\eta_\pm$ is the $\pi$ pulse error on the $|0\rangle \leftrightarrow |\pm\rangle$ transitions.
The spin state is propagated over delay $\tau$ using
\begin{equation}
\mathbf{P}(\tau) = \left(\begin{array}{clcr}
p_{--}(\tau)&p_{0-}(\tau)&p_{+-}(\tau)\\
p_{-0}(\tau)&p_{00}(\tau)&p_{+0}(\tau)\\
p_{-+}(\tau)&p_{0+}(\tau)&p_{++}(\tau)\\
\end{array}\right)
\end{equation} 
according to the equations of motion (Eq.~\ref{eq:matrix_pop_decay}), and
\begin{eqnarray}
\mathbf{c}=\left(\begin{array}{c}
f_0 (1-C)\\
f_0\\
f_0 (1-C)
\end{array}\right)
\end{eqnarray}
accounts for the finite readout contrast (where $f_0$ is the expected photon counts for one readout from state $|0\rangle$, and $C$ is the fluorescence contrast). Finally, the trailing term in (Eq.~\ref{eq:Mmodel}), $b(\tau)$ is a potentially delay-dependent background fluorescence that does not depend on which spin state we initialize or measure.

Note the assumptions underlying this model. First, we assume the spin polarizes equally into $|\pm \rangle$; this will fail in the presence of very fast, unequal relaxation rates comparable to the optical pumping rate. Second, we assume the two states $|\pm\rangle$ exhibit the same fluorescence; this could fail under the same circumstances. Third, we assume the $\pi$ pulses drive only one transition, with no crosstalk (this will fail for small magnetic field). Fourth, we assume the time required for relaxation out of the singlet states (following optical pumping) is negligible relative to relevant decay times. Finally, we assume that the spin dephasing time is negligibly short such that evolution is well described by the rate equation model. Ultimately, we expect this model to be valid for decay rates much slower than the microsecond photophysics and spin dephasing times. 

This model for a generic signal acquisition allows us to calculate an expected value of our measurements. The rate equation solutions $p_{ij}(\tau)$ have some symmetry properties and constraints, specifically $p_{ij}(\tau) = p_{ji}(\tau)$ and $\sum_i p_{ij}(\tau) = \sum_j p_{ij}(\tau) = 1$. Combined with Eq.~\ref{eq:Mmodel}, one can show that the difference
\begin{eqnarray}
    \overline{ S_{00}(\tau)} - \overline{ S_{\pm 0}(\tau)} =    ~~~~~~~~~~~~~~~~~~~~~~~~~~~~~~~~~\nonumber\\
 R  C f_0\frac{ (3\alpha-1)}{2} (1-\eta_\pm)\left(p_{00}(\tau) - p_{\pm 0}(\tau)\right) \label{eq:numdi},
\end{eqnarray}
such that dividing by  $\overline{S_{00}(0)} - \overline{S_{\pm 0}(0)}$ yields a measurement (Eq.~\ref{eq:M}) that is essentially independent of $f_0, C, \alpha, \eta_\pm,$ and $b(\tau)$. Other choices of difference measurements $S_{ij}(\tau) - S_{kl}(\tau)$ yield expressions where the $\pi$ pulse fidelities do not factor out from the population dynamics, but the other parameters do. For example, the most sensitive pulse sequence (see Appendix~\ref{app:comparing-protocols}) involves
\begin{widetext}
\begin{equation}
    \overline{S_{\pm\pm}(\tau)} - \overline{S_{\pm 0}(\tau)} = R C f_0 \frac{ (3\alpha-1) }{2} (1-\eta_\pm)\big[p_{\pm\pm}(\tau) - p_{\pm 0}(\tau)\nonumber - \eta_\pm(p_{00}(\tau)- 2 p_{\pm 0}(\tau) + p_{\pm \pm}(\tau))\big]\nonumber\label{eq:numbest},
\end{equation}
which yields the following pair of normalized measurements
\begin{equation}
\frac{\overline{S_{\pm\pm}(\tau)} - \overline{S_{\pm 0}(\tau)}}{\overline{S_{\pm\pm}(0)} - \overline{S_{\pm 0}(0)}}=e^{-\tau(\Gamma_{+}+\Gamma_{-})}\left\{ \cosh(\tau\mathcal{G})+\frac{(\Gamma_{\pm}-\Gamma_{\mp} + \eta_{\pm}(\Gamma_{\mp}-2\Gamma_{\pm}))\sinh(\tau\mathcal{G})}{(2\eta_{\pm}-1)\mathcal{G}}\right\}  \nonumber \label{eq:fbest}
\end{equation}
\end{widetext}

The inquisitive reader may wonder if it is valid to use the same values of $C, \alpha, f_0$ and $\eta_\pm$ for different signals. In doing so, we are implicitly assuming that they do not vary in between the times when the two signals are measured. For this reason, we employ the pulse sequence of Fig.~\ref{fig:Suggested-measurement-of}(a) that interleaves acquisition of signals $S_{ij}(\tau)$ and repeats the entire pattern $R$ times (rather than first measuring $R$ repetitions to obtain $S_{00}(\tau)$, say, then $R$ repetitions for $S_{+0}(\tau)$, and so on).  This act of interleaving is essential for parameter cancellation when summing over $R\gg 1$ repetitions: expressions like Eq.~\ref{eq:numdi} generalize to a sum over $R$ drifting parameters rather than a simple multiplication by $R$; however, since $p_{ij}(\tau)$ is deterministic, this sum is cancelled by a nearly identical sum in the denominator of Eq.~\ref{eq:M}, provided all 4 ($\approx$100-$\upmu$s-long) signals are acquired each repetition, i.e., before the parameters have time to drift.

\section{Estimating cost and optimal delays\label{app:calculating-cost}}
As discussed in the main text, the cost function (Eq.~\ref{eq:cost})
\begin{equation}
\mathcal{C}(\tau_+,\tau_-)=\sqrt{\left(\frac{\sigma_{\Gamma_{+}}}{\Gamma_{+}}\right)^2 +\left(\frac{\sigma_{\Gamma_{-}}}{\Gamma_{-}}\right)^2}\sqrt{T}.\label{eq:cost_appendix}
\end{equation}
serves as a proxy for sensitivity. In this section, we provide a simple means of estimating $\mathcal{C}$ at the optimal delays $\tau_\pm$, for any of the possible four-signal measurement protocols, in the situation where the true rates $\Gamma_\pm^{\mathrm{true}}$ are known -- this will allow us to compare the sensitivity of different protocols. In the interest of notational continuity, we denote the two measurements as $M_+(\tau_+)$ and $M_-(\tau_-)$, with the understanding that the $\pm$ subscripts indicate the two measurements of the chosen protocol. 

Our first step is to relate the rate uncertainties $\sigma_{\Gamma_\pm}$ after a single measurement pair $M_+$ and $M_-$ to the uncertainties in those measurements ($\sigma_{M_+}$ and $\sigma_{M_-}$ respectively). 
With a flat prior, the posterior distribution after the measurements are performed (Eq.~\ref{eq:bayesian-update}) simplifies to
\begin{equation}\label{eq:approximate-posterior}
\begin{split}
\text{P}(\Gamma_+, \Gamma_-|M_+,M_-) \propto 
e^{-\chi_{+}^{2}-\chi_{-}^{2}},
\end{split}
\end{equation}
where
\begin{align}
\chi_{\pm} \equiv \frac{M_\pm-\tilde{M}_{\pm}(\tau_{\pm},\Gamma_{+},\Gamma_{-})}{\sqrt{2}\sigma_{M_{\pm}}},\label{eq:chi_signals}
\end{align}
and we have assumed sufficient counting statistics that the distributions of $M_\pm$ are approximately Gaussian (see Appendix \ref{app:nonlinear-error}). The standard deviation of $\text{P}$, which is not Gaussian due to the model function $\tilde{M}$, provides a numerical estimate of $\sigma_{\Gamma_\pm}$ for Eq.~\ref{eq:cost_appendix}. However, with sufficiently high signal, we can approximate it as Gaussian by Taylor expanding the exponent in the quantities $\Delta \Gamma_\pm \equiv \Gamma_\pm -\Gamma_\pm^\text{true}$, where $\Gamma_\pm^{\mathrm{true}}$ are the ``true'' values of $\Gamma_\pm$ we hope to measure, which we assume approximately coincide with the location of the maximum of the Gaussian. Specifically,
\begin{align}
    &\chi_{+}^{2}+\chi_{-}^{2}\approx \frac{1}{2}\left[a_+\Delta\Gamma_+^2+a_-\Delta\Gamma_-^2 + 2a_0\Delta\Gamma_-\Delta\Gamma_+\right]
\end{align}
where
%
\begin{eqnarray}
a_\pm &=& \frac{1}{\sigma^2_{M_-}}\left(\frac{\partial \tilde{M}_-}{\partial \Gamma_{\pm}}\right)^2+\frac{1}{\sigma^2_{M_+}}\left(\frac{\partial \tilde{M}_+}{\partial \Gamma_{\pm}}\right)^2\bigg|_{\Gamma_\pm =\Gamma_\pm^{\mathrm{true}}}\\
a_0 &=& \frac{1}{\sigma^2_{M_-}}\frac{\partial \tilde{M}_-}{\partial \Gamma_-}\frac{\partial \tilde{M}_-}{\partial\Gamma_+}+\frac{1}{\sigma^2_{M_+}}\frac{\partial \tilde{M}_+}{\partial \Gamma_-}\frac{\partial \tilde{M}_+}{\partial\Gamma_+}\bigg|_{\Gamma_\pm =\Gamma_\pm^{\mathrm{true}}}.
\end{eqnarray}
The uncertainties and covariance can then be calculated analytically from the resulting Gaussian distributions as
\begin{eqnarray}
\sigma^2_{\Gamma_\pm} =\frac{ a_\mp }{a_+a_- - a_0^2} \text{~~and}~~
\sigma^2_{\Gamma_+\Gamma_-} = \frac{ - a_0 }{ a_+a_- - a_0^2 },
\label{eq:app-analytical_egamma}
\end{eqnarray}
respectively. Note that these uncertainties are now analytic functions of $\tau_\pm$, $\Gamma_{\pm}^{\mathrm{true}}$, and $\sigma_{M_\pm}$. 

We note that the same expressions for $\sigma_{\Gamma_\pm}^2$ can be obtained much more compactly, though perhaps without the same transparency of approximations, by employing the Jacobian matrix
\begin{equation}
    \mathbf{J} = \left(\begin{array}{clcr}
    \frac{\partial \tilde{M}_-}{\partial \Gamma_-}& \frac{\partial \tilde{M}_-}{\partial \Gamma_+}\\
    \frac{\partial \tilde{M}_+}{\partial \Gamma_-}& \frac{\partial \tilde{M}_+}{\partial \Gamma_+}
    \end{array}\right)\bigg|_{\Gamma_\pm =\Gamma_\pm^{\mathrm{true}}}.
\end{equation}
Assuming that there is no correlation between the measurements $M_+$ and $M_-$, the covariance matrix for $\Gamma_\pm$ can be found via
\begin{equation}\label{jacobian}
    \left(\begin{array}{clcr}
   \sigma_{\Gamma_-}^2 & \sigma_{\Gamma_+\Gamma_-}\\
     \sigma_{\Gamma_+\Gamma_-} & \sigma_{\Gamma_+}^2 
    \end{array}\right) 
    = \mathbf{J}^{-1}\cdot
     \left(\begin{array}{clcr}
   \sigma_{M_-}^2 & 0\\
     0 & \sigma_{M_+}^2 
    \end{array}\right)
    \cdot (\mathbf{J}^\intercal)^{-1},
\end{equation}
where $^\intercal$ indicates transpose. This formalism permits an easy extension of the NOB approach to systems with more than two measurements and rates.

Regardless of how $\sigma_{\Gamma_\pm}$ are obtained, substituting $\sigma_{\Gamma_\pm}$ into Eq.~\ref{eq:cost_appendix} and setting $\Gamma_\pm = \Gamma_\pm^{\mathrm{true}}$ and $T = 2 R (\tau_+ +\tau_-) + T_0$  yields an expression for the cost.
The optimal delays $\tau_\pm^\text{opt}$ and minimal cost $\mathcal{C}(\tau_\pm^\text{opt})$ thus depend on which measurement protocol ($\tilde{M}_\pm$) is employed as well as $\Gamma_\pm^{\mathrm{true}}$, $R, T_0$, and the measurement uncertainties $\sigma_{M_\pm}$. We use $\tau_\pm$ and $\mathcal{C}$ in two different ways, for which we evaluate these quantities differently. First, when comparing the minimal cost of different measurement protocols, we assume known $\Gamma_\pm^{\mathrm{true}}$, set $T_0 = 0$ (such that $R$ factors out), and model the measurement uncertainties by nonlinear error propagation (see Appendix~\ref{app:nonlinear-error}) on shot-noise-limited signals (modeled according to  Appendix~\ref{app:Photoluminescence-to-Signal} using specified parameters). This careful modeling allows us to predict as accurately as possible which measurement protocol is most sensitive. 

When finding the optimal delays in the NOB adaptive protocol, we can afford to make more approximations to speed up the computation, since they will not impact the accuracy of our estimates for $\Gamma_\pm$. In this situation (where the true rates are unknown), we employ the mean value $\bar{\Gamma}_\pm$ from the latest distribution to estimate the true rates, use the overhead from the previous measurement for $T_0$,  and set $R$ equal to its target value (in practice $R$ fluctuates by $\approx0.01~\%$ due to the data acquisition implementation). Finally, we approximate equal, delay-independent measurement uncertainties $\sigma_{M_+}\approx \sigma_{M_-}\approx \sigma_M$; $\sigma_M$ then factors out. We thus estimate the optimal delays by minimizing the following function, where the delay-dependence enters through the form of $T$ and $\tilde{M_\pm}$:
\begin{widetext}
\begin{equation} \label{eq:approxC}
    \frac{\mathcal{C}}{\sigma_M}\approx \frac{\sqrt{T}}{\Gamma_- \Gamma_+}\times\\\sqrt{
    \frac{\left(\Gamma_-\frac{\partial \tilde{M}_-}{\partial\Gamma_-}\right)^2 +\left(\Gamma_-\frac{\partial \tilde{M}_+}{\partial\Gamma_-}\right)^2 +\left(\Gamma_+\frac{\partial \tilde{M}_-}{\partial\Gamma_+}\right)^2 +\left(\Gamma_+ \frac{\partial \tilde{M}_+}{\partial\Gamma_+} \right)^2}
    {\left(\frac{\partial \tilde{M}_+}{\partial\Gamma_-}\frac{\partial \tilde{M}_-}{\partial\Gamma_+}- \frac{\partial \tilde{M}_-}{\partial\Gamma_-}\frac{\partial \tilde{M}_+}{\partial\Gamma_+}\right)^2}
    }.
\end{equation}
\end{widetext}

It is worth noting that this approach could in principle be applied to measurements beyond the ones considered here (not restricted to the NV model of Appendices A-B). $\tilde{M}_\pm$ could represent a more general parametric model for an experiment, from which more general parameters $\Gamma_\pm$ are extracted, and the cost function could be optimized with respect to any controllable measurement parameter that appears in the model. Moreover, Eq.~\ref{jacobian} and Eq.~\ref{eq:cost_appendix} are readily generalized to larger numbers of measurements and extracted parameters, providing a framework for optimizing multi-parameter measurements. 

\section{Comparing different protocols}\label{app:comparing-protocols}

\begin{figure}
\includegraphics[width=8.6cm]{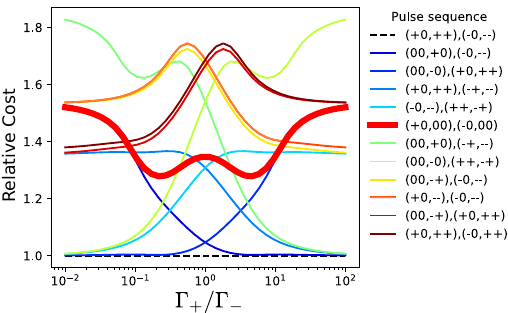}\caption{\label{fig:Rank-verus-rates}
Cost function $\mathcal{C}$ for the 12 lowest-cost measurement pairs, evaluated at optimal delays $\tau^{\text{opt}}_\pm$ over a range of ratios $\Gamma_+/\Gamma_-$, all scaled by the $(+0,++),(-0,--)$ protocol for comparison. In these calculations, we assume $|0\rangle$-state counts $f_0=0.02$, contrast $C=0.24$, pump fidelity $\alpha=1$, $\pi$-pulse error $\eta_{\pm}=0$, and overhead time $T_0 = 0$ (which eliminates dependence on $R$). We do not model any drift effects.
The red thick curve, $(+0,00),(-0,00)$, is the drift-insensitive protocol employed in the main text.}
\end{figure} 

Since each of the 8 indices of the measurement pairs $(ij,kl), (pq,rs)$ can take on one of 3 values $\{0,+,-\}$, there exist $3^8=6561$ possible 4-signal measurement protocols, 36 of which are independent and have non-zero expected denominator ($\bar{S}_1(0) - \bar{S}_2(0) \ne 0$). To ensure we have chosen the optimal protocol, we use the cost function $\mathcal{C}$ (Eq.~\ref{eq:cost_appendix}) to compare their relative performances. For this, we use an overhead-free ($T_0=0$) total time $T = 2 R (\tau_+ + \tau_-)$, assumed rates $\Gamma_\pm$, and the analytic forms for $\sigma_{\Gamma_\pm}$ (Eq.~\ref{eq:app-analytical_egamma}) evaluated with the predicted uncertainty in $\sigma_{M_\pm}$ (Eq.~\ref{eq:app:signa_estimation_fromAandZ} from Appendix \ref{app:nonlinear-error} below). Figure \ref{fig:Rank-verus-rates} shows the cost $\mathcal{C}$ for the 12 best measurement pairs, evaluated at optimal delays $\tau^{\text{opt}}_\pm$ over a range of ratios $\Gamma_+ / \Gamma_-$, all scaled by the cost of the $(+0,++),(-0,--)$ protocol. Note that rescaling both decay rates $\Gamma_\pm$ will not affect this result, since this scales $\tau_\pm^\text{opt}$ in the same way for all protocols.

Importantly, the cost associated with the measurement pair ($+0,00$),($-0,00$) discussed in the main text (dashed maroon curve) is within $\approx50~\%$ of that of the pair $(+0,++),(-0,--)$, which has the best cost value over the entire range. In our case, the associated loss of sensitivity represents an acceptable trade-off for drift insensitivity. However, $(+0,++)$,$(-0,--)$ may be a better choice in systems with excellent $\pi$ pulse stability, even with drifting laser intensity.

\section{Avoiding bias in measurements
\label{app:nonlinear-error}}

\begin{figure}
\includegraphics[width=8cm]{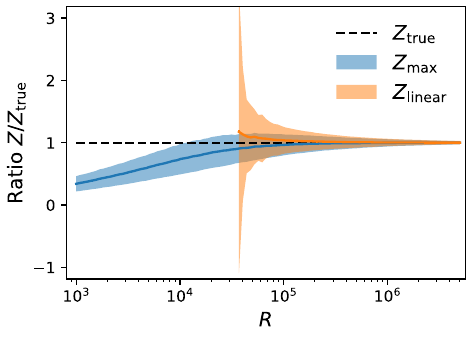}
\caption{\label{fig:bias_estimate}
\textbf{Example Monte Carlo sampling of $Z$,} using the same parameters as in Fig.~\ref{fig2:adaptive-approach} with $S_1 = S_{00}$ and $S_2 =S_{+0}$. Evolution of $Z$ with $R$, relative to the true value $Z_{ \mathrm{true}}$ from the model Eq. \ref{eq:Mmodel}.
The solid blue and orange curves represent the mean of $10^4$ repeated evaluations of $Z$, using the nonlinear \ref{eq:app:nonlinear_formula_1overB} (blue) and linear (orange) error-propagation.
The filled bands indicate the standard deviation of the distribution of outcomes. $Z_{ \mathrm{linear}}$ are missing when at least one value of $\bar{\Delta}=0$ occurred among the $10^4$ samples. }
\end{figure} 

Since a given ($n^\text{th}$) measurement
\begin{equation}
M^{(n)}(\tau) = \frac{S_1^{(n)}(\tau) - S_2^{(n)}(\tau)}{S_1^{(n)}(0)-S_2^{(n)}(0)}    
\label{eq:app-measurement-ratio}
\end{equation}
involves a ratio, we must be wary of measurement bias introduced by the fluctuations in the denominator signals $S_j^{(n)}(0)$. In this section, we describe one simple method to somewhat reduce this bias while avoiding zero-valued denominators, and how to choose a reasonable number of readouts $R$ for the experiment. 

If the individual signals follow a Gaussian distribution, the denominator $\Delta_n \equiv S_1^{(n)}(0)-S_2^{(n)}(0)$ of this measurement will also be drawn from a Gaussian distribution of mean $\bar{\Delta}=\bar{S}_1(0)-\bar{S}_2(0)$ and variance $\sigma_{\Delta}^2 = \sigma_{S_1(0)}^2 + \sigma_{S_2(0)}^2$. However, its reciprocal $Z \equiv 1/\Delta$ follows a non-Gaussian distribution
\begin{equation}
   \text{P}(Z) \propto \frac{1}{Z^2}\exp\{-(1/Z- \bar{\Delta})^2/2\sigma_{\Delta}^2\},  
\end{equation}
since $\text{P}(\Delta) d\Delta = \text{P}(Z)dZ$ for a monotonic function $Z(\Delta)$. 

By choosing the appropriate ordering of $S_1(0)$ and $S_2(0)$ we can ensure $\Delta$'s (and thus $Z$'s) positivity, and approximate $\text{P}(Z)$ as Gaussian around its maximum at $Z_\text{max} > 0$. Specifically, setting $d\log{P(Z)}/dZ = 0$ and Taylor expanding $\log{P(Z)}$ around $Z_\text{max}$ provides a variance $\sigma_Z^2 = (-d^2\log{P(Z)}/dZ^2)^{-1}|_{Z = Z_\text{max}}$, yielding
\begin{eqnarray}
Z_\text{max} & = &\frac{1}{4\sigma_{\Delta}^{2}}\left(\sqrt{\bar{\Delta}^2+8\sigma_{\Delta}^{2}}-\bar{\Delta}\right)\label{eq:app:nonlinear_formula_1overB}\\
\sigma_{Z} & =&
\frac{Z_\text{max}^2 \sigma_\Delta}{\sqrt{2 -  Z_\text{max}\bar{\Delta}}} \label{eq:app:nonlinear_formula_error_1overB}
\end{eqnarray}
We can now construct an improved estimate of the measurement mean
\begin{eqnarray}\label{eq:app:Mbar}
\bar{M}^{(n)} & = & A_n Z_\text{max},
\end{eqnarray}
where $A_n \equiv S_1^{(n)}(\tau) - S_2^{(n)}(\tau)$, and uncertainty
\begin{eqnarray}
\sigma_{M} & =&M^{(n)}\sqrt{\left(\frac{\sigma_{A_n}}{A_n}\right)^{2}+\left(\frac{\sigma_{Z}}{Z_\text{max}}\right)^{2}}\label{eq:app:signa_estimation_fromAandZ},
\end{eqnarray}
where $\sigma_A^2 = \sigma_{S_1(\tau)}^2 + \sigma_{S_2(\tau)}^2$. Note this approach is used to update the posterior in both the adaptive and non-adaptive data analysis.

Equations~\ref{eq:app:nonlinear_formula_1overB}-\ref{eq:app:nonlinear_formula_error_1overB} still represent a biased estimator, and so it is important to explore the signals required to sufficiently suppress the remaining bias. 
To get a sense of scale, we numerically generate $10^4$ instances of $S_1(0)$, $S_2(0)$ (from the model Eq. \ref{eq:Mmodel}) to evaluate $Z$. 
Each of $S_1(0)$ and $S_2(0)$ is taken from a Poissonian distribution with the same parameters as in Fig. \ref{fig2:adaptive-approach}, with different numbers of repetitions $R$.
We compare the average value of $Z$ obtained with our improved estimate (Eq. \ref{eq:app:nonlinear_formula_1overB}) and a linear error propagation ($Z_{ \mathrm{linear}}$) to the underlying true value from the model ($Z_{ \mathrm{true}} = 1/[\frac{1}{2} C f_0 R (3 \alpha -1)(1 - \eta_\pm)]$).
Figure~\ref{fig:bias_estimate} shows how the mean estimate converges to the true value above $R\approx10^5$, motivating our choice of $R=10^6$. This result is corroborated by numerical comparisons of true and extracted rate parameters in Fig.~\ref{fig2:adaptive-approach}(c), where any residual bias in our estimation is smaller than our measurement uncertainty.

\end{document}